\documentstyle[aps,epsf]{revtex}
\newcommand{\bref}[1]{(\ref{#1})}
\newcommand{\vth}{\vartheta} 
\newcommand{\nn}{\nonumber}
\title{Multi-Bunch Solutions of Differential-Difference 
Equation for Traffic Flow}
\author{Ken Nakanishi}
\address{Department of Physics, Nagoya University, Nagoya 464-0814,
Japan}
\date{\today}
\begin{document}
\twocolumn[\hsize\textwidth\columnwidth\hsize\csname
@twocolumnfalse\endcsname
\maketitle
\begin{abstract}
Newell-Whitham type car-following model with hyperbolic tangent optimal 
velocity function in a one-lane circuit has a finite set of the exact
solutions for steady traveling wave, which expressed by elliptic theta
function. Each solution of the set describes a density wave with
definite number of car-bunches in the circuit. By the numerical
simulation, we observe a transition process from a uniform flow to the
one-bunch analytic solution, which seems to be an attractor of the
system. In the process, the system shows a series of cascade
transitions visiting the configurations closely similar to the
higher multi-bunch solutions in the set.
\end{abstract}
\vskip6pt
\hspace*{1.8cm}
\vskip1.9pc]
\narrowtext


\section{Introduction}
Traffic flow in a no-passing freeway has extensively study with the
car-following models. Some of these models can describe the
spontaneous generation of density wave along the road, the traffic
congestion, as the collective motion of cars. The optimal velocity
(OV) model \cite{OVM1,OVM2} reveals the features of the congested
flow. In a certain range of the mean headway, cars gradually compose
bunch of cars, which moves backward with a constant velocity.
The most of the studies are achieved by the numerical simulations
under the cyclic boundary condition, which solve the equation of
motion of the model;
\begin{equation}
{\ddot x}_n(t)=a\left[V(\Delta x_n(t))-{\dot x}_n(t)\right].
\label{OVM}
\end{equation}
$x_n$ and $\Delta x_n$ denote the position of the $n$-th car and 
the headway of the car, $x_{n-1}-x_n$, respectively. Drivers try
to adjust their car's velocity to the optimal velocity
$V(\Delta x)$ according to the headway. The constant $a$ parameterizes
the sensitivity in the adjustment, which should be less than a
critical value to generate the congestion.

In the density wave of the congested flow, high density region, the
bunch of cars, alternates with low density one. The cars run almost
same headway and velocity in each of the regions, and in the interface
of the regions the density changes rapidly exhibiting the
characteristic kink-like shape. The shape of the kink on the interface
seems to be determined only by the sensitivity $a$ and the OV function
$V(\Delta x)$, irrespective of the initial condition or the mean
headway. In the vicinity of the critical point, the interface can be
approximately described by the kink solution of the modified
Korteweg-de~Vries equation \cite{Sasa}. For some particular choices of
the OV function, the exact solution which describes the interface has
been obtained \cite{Sugiyama,Nakanishi}. However, in the numerical
simulations, the generation process of the congestion displays much
complicated aspects. A number of bunches with various length may arise
in the circuit, and the fusion of bunches is often observed. One can
predict neither the number of the bunches nor the lengths of them.

On the other hand, the traditional car-following model
\cite{CFM}, which is described by the first order differential
difference equation,
\begin{equation}
{\dot x}_n(t+\tau)=V(\Delta x_n(t)),
\label{SCFM}
\end{equation}
has been studied in the traffic engineering for a long time. The
reaction time $\tau$ represents the time lag which it takes the car to
respond to the change of motion. Since the OV model \bref{OVM} can be
considered as the truncated Taylor expansion, ${\dot x}_n(t+\tau)
\simeq{\dot x}_n(t)+\tau{\ddot x}_n(t)$, of Eq. \bref{SCFM}, these
models rather resemble each other in the qualitative behavior,
especially in the generation of the steady traveling wave. It is
shown numerically that Eq. \bref{SCFM} with a hyperbolic tangent OV
function has the congested flow solutions quit similar to ones for the
corresponding OV model \cite{Hasebe}. 

Early in the 90's, Whitham \cite{Whitham} showed that the model with
$V(\Delta x)$ given by an exponential function has the pulse-like
exact solution for steady traveling wave described by the elliptic
functions. Although his choice of the function $V(\Delta x)$
admits no existence of the car-bunching solution mentioned above, he
pointed out a crucial relation between the time lag $\tau$ and the
propagating velocity of the traveling wave. (See Eq. \bref{Whitham}.)
We may call it Whitham condition. Recently \cite{IIN}, it is
demonstrated analytically that a class of the car-bunching solutions
represented by the elliptic theta functions exactly satisfies
Eq. \bref{SCFM} with the hyperbolic tangent OV function, 
\begin{equation}
V(\Delta  x)
 = \xi~+~\eta \tanh\left[\frac{\Delta x -\rho}{2\sigma}\right].
\label{hytanh}
\end{equation}
The authors of Ref. \cite{HNS} report that the Whitham condition holds
in the numerical simulation for much wider choice of the OV function.

In this paper, we will investigate the structure of the class of the
exact traveling wave solutions of Eq. \bref{SCFM}. Under the cyclic
boundary condition, each solutions corresponds to a periodic density
wave with definite length and number of bunches. Generally, some
solutions different in the number of bunches are possible. They will
be  mentioned as the multi-bunch solutions. We also perform the
numerical simulation and observe the generation process of the density
wave which shows a relaxation to one of the analytic solutions. 

In the next section, we investigate the parameter space of the exact
solutions and find the finite set of the multi-bunch solutions. In
section III, allowed parameters for the multi-bunch solution are
determined. These parameters will be used in the numerical simulation
in section IV. The final section is devoted to summary and
discussions. Some mathematical formula for the elliptic functions and
the exact solutions are summarized in the appendix.


\section{Multi-bunch solutions in a circuit}
We begin by summarizing our previous results \cite{IIN} of the exact
solution with a width parameter $\delta$:
\begin{equation}
x_n(t)= Ct~-n h~+A~{\ln} \frac
{\vth_0\left(\nu t-\frac{n}{\lambda}-\frac{1}{2\lambda}+\delta,q~\right)}
{\vth_0\left(\nu t-\frac{n}{\lambda}-\frac{1}{2\lambda}-\delta,q~\right)},
\label{sol1}
\end{equation}
where $\vth_0 (v,q)$ is the elliptic theta function and $q$, $A$,
$\lambda$, $\nu$, $\delta$, $C$ and $h$ are ansatz parameters which
characterize the solution. $q$ is modulus parameter of the theta
function. $h$ is the mean headway of $N$ cars in the
circuit with length $L\equiv Nh$. The traffic flow expressed by
\bref{sol1} displays alternate appearance of {\it high density region}
and {\it low density region}. Thus, the traffic in the circuit is
divided into several number of ``bunches'', which move backward.  
The cyclic boundary condition, $x_{n+N}(t)=x_n(t)-Nh$, and the
periodicity of theta function, $\vth_0(v+1,q)=\vth_0(v,q)$, 
imply that $N/\lambda$ must be an integer, which coincides with the
number of bunches $n_{\rm b}$; 
\begin{equation}
\lambda=\frac{N}{n_{\rm b}},\qquad(~n_{\rm b}:~{\rm integer}~).
\label{bunch}
\end{equation}
Then, the ``wavelength'' $\lambda$ is the number of cars
within a successive pair of high and low density region, approximately.
The width parameter $\delta$, which ranges in $0<2\delta<1$,
determines the ratio of low density region in a wavelength. 
$2\delta\lambda$ and $(1-2\delta)\lambda$ are the approximate number of
cars in the low and high density regions, respectively.

The exact solution \bref{sol1} satisfies the equation of motion
\bref{SCFM} with the OV function \bref{hytanh}. As discussed in
\cite{IIN}, the traffic model with given time lag constant $\tau$ can
admit the exact solution \bref{sol1} only if the Whitham condition
\begin{equation}
2\nu\lambda\tau=1,
\label{Whitham}
\end{equation}
is met. Under the condition, we find out four relations between the
ansatz parameters and the coefficients of the equation of motion,
$\xi$, $\eta$, $\rho$ and $\sigma$, as stated in the appendix
\bref{a:xi1}--\bref{a:sigma}. Among the seven ansatz parameters, the
mean headway $h$ is already determined by the system size $L$ and
$N$. Furthermore, by the Whitham condition, we find that $\nu$ is not an
independent parameter but proportional to the inverse of
$\lambda$. The remaining five parameters $q$, $A$, $\lambda$, $\delta$
and $C$ are subject to the above four relations. 
It may seem that a parameter remains free and the system have a
one-parameter family of the exact solutions. However, we notice that
$\lambda$ is discretized as \bref{bunch} so that the bunch number
$n_{\rm b}$ is an integer. Moreover, the number of the bunches can not
exceed the total number of the cars $N$ at least. (We will see that
much stronger restriction exists for the maximum number of the bunches
later.) Consequently, the ansatz \bref{sol1} gives us the {\it finite}
set of the exact solutions which correspond to the $n_{\rm b}$-bunch
states of traffic congestion. We call them the ``multi-bunch
solutions''.

To examine the multi-bunch solutions, let us precisely analyze the
four relations \bref{a:xi1}--\bref{a:sigma};
\begin{eqnarray}
\xi&=&C+\frac{A\beta}{2\tau}\frac{d}{d\beta}
\ln\frac{\vth_1(2\delta+\beta)}{\vth_1(2\delta-\beta)},
\label{xi}\\
\eta&=&\frac{A\beta}{2\tau}\frac{d}{d\beta}\ln
\frac{\vth_1^2(\beta)}{\vth_1(2\delta+\beta)\vth_1(2\delta-\beta)},
\label{eta}\\
\rho&=&h-A\ln\frac{\vth_1(2\delta-\beta)}{\vth_1(2\delta+\beta)},
\label{rho}\\
\sigma&=&A.
\label{sigma}
\end{eqnarray}
Here and hereafter, $\lambda$ is changed to a much convenient
variable, $\beta$, which defined in the appendix as
\begin{equation}
\beta=\frac{1}{2\lambda}.
\end{equation}
Since the variable is also equal to $n_{\rm b}/(2N)$, it will be
mentioned as the ''bunch parameter''. (Note that, in the expressions
\bref{xi} and \bref{eta}, $\beta/\tau$ is substituted for $\nu$ by
using the Whitham condition.) The equation \bref{sigma} simply says that
$A$ is identical to $\sigma$. From the equation \bref{xi}, we realize
that the velocity parameter $C$ can be expressed as an function of
$q$, $\delta$ and $\beta$ (or $\lambda$). Thus, the two nontrivial
relations \bref{eta} and \bref{rho} should be solved to establish the
existence of the multi-bunch solutions and to find out the allowed
region of the parameters.

First, we treat the equation \bref{eta}, whose variables in the theta
function can be decomposed by using the addition formula as shown in
the appendix. Replacing $\vth_1(v)/\vth_0(v)$ by the Jacobi's
elliptic function $\sqrt{k}~{\rm sn}2Kv$ in \bref{a:eta2}, it becomes
\begin{equation}
\frac{\tau}{\tau_c}=-\frac{\beta}{2}\frac{d}{d\beta}
\ln\left[\frac{1}{{\rm sn}^22K\beta}-\frac{1}{{\rm sn}^24K\delta}\right],
\label{tau_cond}
\end{equation}
where $A/\eta~(=\sigma/\eta)$ is denoted by $\tau_c$ because of the
following reason. A linear analysis \cite{Whitham,Herman} gives the
instability condition for a uniform flow described by
$x_n^{(0)}(t)=V(h)t~-n h$,
\begin{equation}
2\tau V'(h)\equiv \tau\frac{\eta}{\sigma}
{\rm sech}^2 \left[\frac{h-\rho}{2\sigma}\right]
>\frac{\pi/N}{\sin\pi/N}. 
\label{linear}
\end{equation}
(Note that the right hand side of \bref{linear} is almost equal to
one, providing that $N$ is not too small.) It follows that
$\eta/\sigma$, the maximum value of $2 V'(h)$, determines the minimum
value of time lag, $\tau_c$, for which the uniform flow becomes
linearly unstable. One expects that when the condition,
\begin{equation}
\tau > \tau_c \equiv \frac{\sigma}{\eta},
\label{linear2}
\end{equation}
is satisfied, there exist the uniform flow which decays to develop
a congested flow. The $\tau_c$ may be referred to as the critical
time lag. By performing $\beta$-derivative, the equation
\bref{tau_cond} can be easily solved with respect to
${\rm sn}4K\delta$ and gives 
\begin{equation}
{\rm sn}^24K\delta=\frac{{\rm sn}^22K\beta}
{{\displaystyle 1-\frac{\tau_c}{\tau}
\frac{2K\beta~{\rm cn}2K\beta~{\rm dn}2K\beta}{{\rm sn}2K\beta}}}.
\label{delta1}
\end{equation}
Since ${\rm sn}^24K\delta\leq 1$, the right hand side of the above
equation should not exceed 1. Thus, we obtain the solvability
condition, which guarantees the existence of $\delta$, as
\begin{eqnarray}
\frac{\tau_c}{\tau}\leq
\frac{{\rm sn}2K\beta~{\rm cn}2K\beta}{2K\beta~{\rm dn}2K\beta}.
\label{solvcon}
\end{eqnarray}

\begin{figure}
\leavevmode
\epsfxsize=8cm
\setlength{\unitlength}{1cm}
\begin{picture}(8,8.5)
\put(-0.4,1.3){\epsfbox{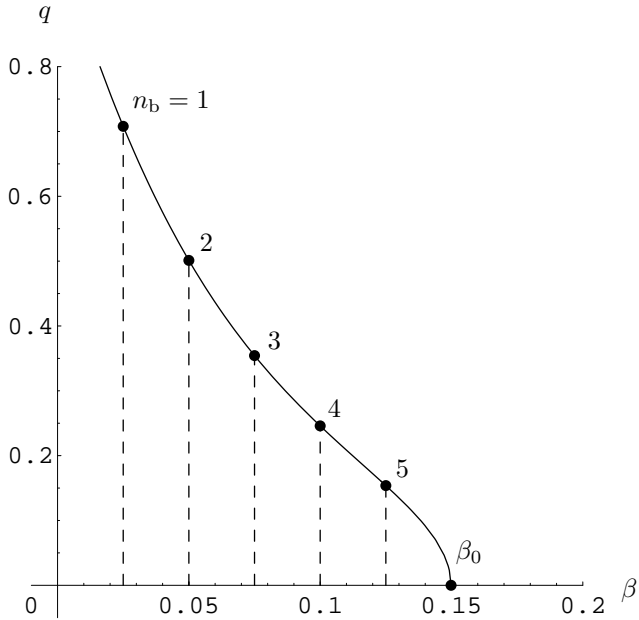}} 
\put(7.8,0.5){$\beta$}
\put(5.6,1){$\beta_0$}
\put(0.05,8.2){$q$}
\put(1.3,7){$n_{\rm b}=1$}
\put(2.2,5.1){$2$}
\put(3.1,3.8){$3$}
\put(3.9,2.9){$4$}
\put(4.8,2.1){$5$}
\end{picture}
\caption{The allowed region of the solvability condition for
$\tau_c/\tau=0.85869$. The maximum value of bunch parameter $\beta_0$
is 0.149835. The vertical dashed lines indicate the ranges of $q$ for
some possible $\beta$ with $N=20$, which correspond to
$n_{\rm b}=1,\ 2,\ 3,\ 4,\ 5$.}
\end{figure}

The inequality gives us the restriction for the number of bunches for
given $\tau$. The right hand side of it only depends on $\beta$ and
$q$, where the complete elliptic integral of the first kind, $K$, and
the modulus of Jacobi's elliptic functions, $k$, are given in
\bref{a:jacobi} as the function of $q$. In FIG. 1, we show the allowed
region of \bref{solvcon} in the $(\beta,\,q)$-plane for
$\tau_c/\tau=0.85869$. Note here that $\beta(=n_{\rm b}/(2N))$ ranges in
$0<\beta<1/2$, since $n_{\rm b}<N$. It can be easily checked that the
allowed region actually disappears for $\tau<\tau_c$, which agrees
with the linear analysis \bref{linear2}. The boundary curve in FIG. 1
crosses the $\beta$-axis at $\beta_0$, the maximum value of
$\beta$. It is given by a solution of the equation obtained in the 
$q\to 0 ~(k\to0,\ K\to\pi/2)$ limit of the equality of \bref{solvcon}:
\begin{eqnarray}
\frac{\tau_c}{\tau}=\lim_{q \to 0}
\frac{{\rm sn}2K\beta_0~{\rm cn}2K\beta_0}{2K\beta_0~{\rm dn}2K\beta_0}=
\frac{\sin2\pi\beta_0}{2\pi\beta_0}.
\label{beta_max} 
\end{eqnarray}
Thus, $\beta_0$ is determined only by the value of
$\tau_c/\tau$. The existence of the nontrivial upper bound of $\beta$
implies that the number of the bunches is also restricted much more
than $n_{\rm b}<N$. The maximum value of the number of the bunches
$n_{\rm b}^{\rm max}$ is given by
\begin{eqnarray}
n_{\rm b}^{\rm max}=[2N\beta_0],
\label{bunch_max}
\end{eqnarray}
where $[x]$ is the maximal integer that does not exceed $x$. 
Consequently, when the time lag $\tau$ and the total car number $N$
are given, the system has $n_{\rm b}^{\rm max}$ exact multi-bunch
solutions with the bunch parameters
\begin{equation}
\beta=
\frac{1}{2N},~\frac{2}{2N},~\dots,~\frac{n_{\rm b}}{2N},
~\dots,~\frac{n_{\rm b}^{\rm max}}{2N}.
\label{beta's}
\end{equation}

When we specify one of the possible bunch numbers or $\beta$'s, the
solvability condition \bref{solvcon} gives us an allowed range of
modulus parameter $q$,
\begin{equation}
0\leq q\leq q_{\rm max},
\label{q_max}
\end{equation}
where $q_{\rm max}$ is the value such that the equality of the
solvability condition is held for the selected $\beta$. The vertical
dashed lines in FIG. 1 show the ranges of $q$ for some possible
$\beta$ with $N=20$, where $n_{\rm b}^{\rm max}=5$. 


\section{Construction of the multi-bunch solution}
In this section, we determine the width parameter $\delta$ and the
modulus parameter $q$ for any possible $n_{\rm b}$ (or $\beta$) to
construct the $n_{\rm b}$-bunch solution. In the allowed range
\bref{q_max}, the equation \bref{delta1} has two branches of $\delta$
as functions of $q$. One, which stays in $0<2\delta_-<1/2$, is given
by
\begin{eqnarray}
2\delta_-(q)=\frac{1}{2 K}{\rm sn}^{-1}\frac{{\rm sn}2K\beta}
{\sqrt{\displaystyle 1-\frac{\tau_c}{\tau}
\frac{2K\beta~{\rm cn}2K\beta~{\rm dn}2K\beta}{{\rm sn}2K\beta}}}, 
\label{delta2}
\end{eqnarray}
where we take the branch $0<{\rm sn}^{-1}<K$ for the inverse Jacobi's
function ${\rm sn}^{-1}$. The other branch $1/2<2\delta_+<1$,
which corresponds to $K<{\rm sn}^{-1}<2K$, is obtained as
$2\delta_+(q)\equiv1-2\delta_-(q)$. These two branches are connected
each other at $q=q_{\rm max}$ with $2\delta=1/2$ as depicted in FIG. 2.
On $2\delta=1/2$, the low and high density regions in the traffic flow 
contains same number of cars. Eq. \bref{rho} tells us that such a
``symmetric flow'' can be yielded when the mean headway $h$ is
equal to $\rho$, the inflection point of the OV function, since 
$\vth_{1}(1/2+\beta)=\vth_{1}(1/2-\beta)$ regardless of $\beta$ or $q$.
As long as $2\delta\ne1/2$, the traffic flow becomes asymmetric. For
the first branch $0<2\delta_-<1/2$, the low density region contains
more cars than high density one and the mean headway $h$ is less than
$\rho$. Contrary, for the second branch $1/2<2\delta_+<1$, the
congested region has more cars and $h$ exceeds $\rho$. 

\begin{figure}
\leavevmode
\epsfxsize=8cm
\setlength{\unitlength}{1cm}
\begin{picture}(8,6)
\put(-0.4,0.1){\epsfbox{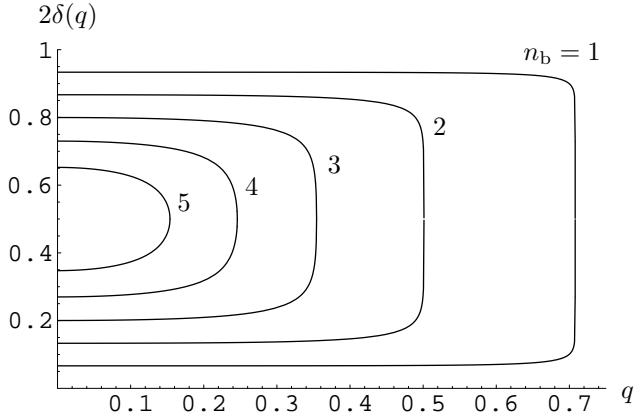}} 
\put(7.8,0.5){$q$}
\put(0.05,5.5){$2\delta(q)$}
\put(6.5,5){$n_{\rm b}=1$}
\put(5.3,4){$2$}
\put(3.9,3.5){$3$}
\put(2.8,3.2){$4$}
\put(1.9,3){$5$}
\end{picture}
\caption{The width parameter $2\delta_\pm(q)$ for
$\tau_c/\tau=0.85869$ with $N=20$. Each curve corresponds to the possible
bunch number $n_{\rm b}$
in FIG. 1.}
\end{figure}

By plugging $2\delta_\pm(q)$ into \bref{rho}, we obtain the relation
$h$ and between the modulus parameter $q$;
\begin{equation}
h_\pm(q)=\rho+\sigma\ln
\frac{\vth_1(2\delta_\pm(q)-\beta,q)}{\vth_1(2\delta_\pm(q)+\beta,q)},
\label{h(q)}
\end{equation}
where $h_-(<\!\!\rho)$ and $h_+(>\!\!\rho)$ correspond to the two
branches of $\delta$. (Note here that $h_-(q)+h_+(q)=2\rho$.) The
combined entire function $h(q)$ is shown in FIG. 3. We can determine
the modulus parameter $q$, within a certain range of mean headway
$L/N$, by solving $h(q)=L/N$, and obtain a exact multi-bunch solution
of \bref{SCFM} by calculating $\delta$ and other parameters through
the equations above.

In some case, we may find several solutions for $q$, which correspond
to same or different bunch number. Although the exact solution can
construct for each $q$, the stability of these solutions is other
problem. We will discuss the issue in another work\cite{bifur}. In
FIG. 3, the dashed lines indicate the linearly unstable region
\bref{linear} of the (common) headway of the uniform flow
$x_n^{(0)}(t)$;
\begin{equation}
|h-\rho|<2\sigma{\rm Arccosh}
\sqrt{\frac{\tau}{\tau_c}\frac{\sin\pi/N}{\pi/N}}.
\label{unstable}
\end{equation} 
Within the range, we may expect that the unstable uniform flow will
grow into one of the analytic solutions. However, the generation
process of the multi-bunch solution from a given initial configuration
and how one of the possible analytic solutions is selected may be more
complicated problem.

\begin{figure}
\leavevmode
\epsfxsize=8cm
\setlength{\unitlength}{1cm}
\begin{picture}(8,5.5)
\put(-0.4,0.1){\epsfbox{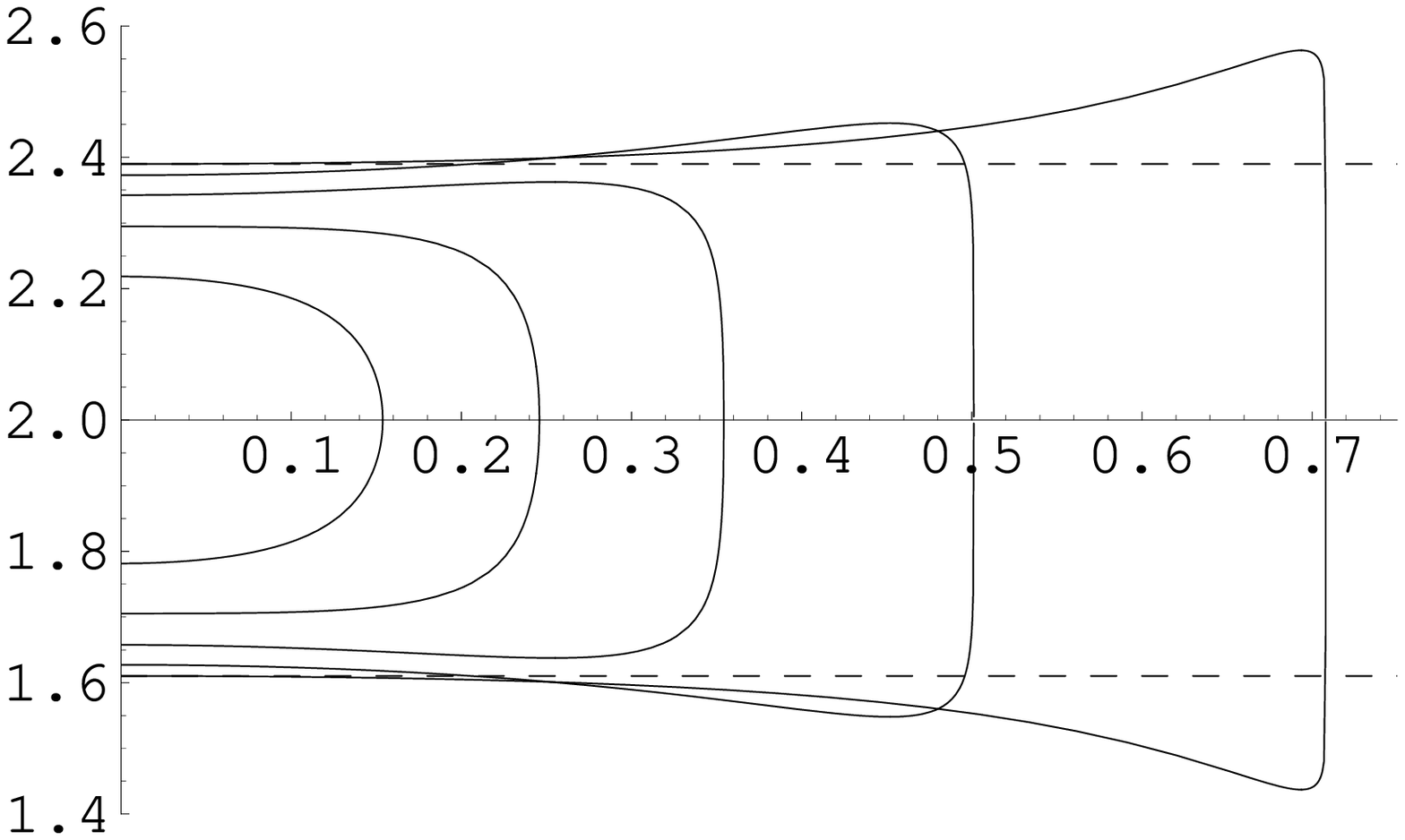}} 
\put(7.8,2.65){$q$}
\put(0.05,5.4){$h(q)$}
\put(6.5,5){$n_{\rm b}=1$}
\put(5.3,3.9){$2$}
\put(3.9,3.5){$3$}
\put(2.8,3.2){$4$}
\put(1.9,3){$5$}
\end{picture}
\caption{The mean headway $h_\pm(q)$ for $\tau_c/\tau=0.85869$,
$\rho=2$, $\sigma=1/2$ with $N=20$. Each curve corresponds to the possible
bunch number $n_{\rm b}$ in FIG. 1 and 2. The horizontal dashed lines
indicate the critical values of headway for the linear instability,
which coincide with the $q\rightarrow0$ limits of $h(q)$ for $n_{\rm b}=1$.}
\end{figure}


\section{Numerical Simulations}
To investigate the generation process of the multi-bunch solution from 
a uniform flow, we perform numerical simulations by solving the
differential-difference equation \bref{SCFM}. We adopt an OV function
\begin{equation}
V(\Delta x)=\tanh(\Delta x-2)+\tanh2,
\label{sym-OV}
\end{equation}
where the coefficients are chosen as $\xi=\tanh2$, $\eta=1$, $\rho=2$
and $\sigma=0.5$ ($\tau_c=0.5$). The time lag $\tau=0.58828$ gives 
$\tau_c/\tau=0.85869<1$. We prepare 20 cars and arrange them so that
they form a uniform flow with a common headway $h=1.88571$ and an
initial velocity $V(h)=0.850233$ in a circuit whose length is
$L=37.7142$. The uniform flow is sustained for a duration of $\tau$ to
prepare the initial function of the differential-difference
equation. The solvability condition \bref{solvcon} tells us that
maximally five-bunch mode is allowed. The modulus parameters $q$ for
the possible bunch numbers $n_{\rm b}=1,\ 2,\ 3,\ 4,\ 5$ are
0.70792140328755, 0.50113376, 0.3536167, 0.2418044, 0.140292,
respectively. Since the uniform flow is linearly unstable, it starts
developing the density wave. In this case, the one-bunch mode is
generated after enough relaxation time, $t\simeq6\times10^4$,
ultimately. The result, which is displayed in FIG. 4 by dots, agrees
quite well with the analytic one-bunch solution with
$q=0.70792140328755$ as shown by the thin line in the figure.

\begin{figure}
\leavevmode
\epsfxsize=8cm
\setlength{\unitlength}{1cm}
\begin{picture}(8,5.5)
\put(-0.4,0.1){\epsfbox{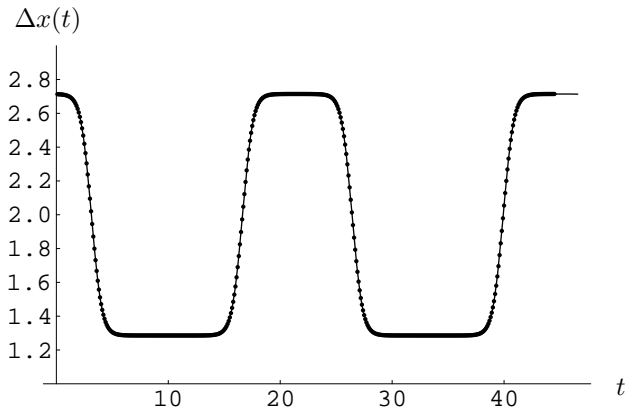}} 
\put(7.8,0.4){$t$}
\put(-0.2,5.3){$\Delta x(t)$}
\end{picture}
\caption{The result of a numerical simulation with $N=20$,
$\tau_c/\tau=0.85869$, $V(\Delta x)=\tanh(\Delta x-2)+\tanh2$ after
 Dots show the headway of a car.
The line shows the 
corresponding one-bunch solution with $q=0.70792140328755$.}
\end{figure}

We observe much interesting phenomena in the generation process of the
analytic exact solution. At first, until $t\simeq300$, the initial
uniform flow grows into a three-bunch configuration which closely
resembles the exact three-bunch solution. Although the configuration
keeps its shape for a while, it is gradually distorting and one of the 
bunches get closer to the other. At $t\simeq4680$, the fusion of
bunches occurs and it is transformed into a two-bunch configuration.
The two-bunch configuration lives about ten times longer than the
three-bunch one. Subsequently, a bunch starts shrinking. The bunch is
absorbed by the other at $t\simeq51560$ as shown in Fig. 5, and then
the exact one-bunch solution is finally accomplished. The simulation
has been continued until $t=2\times10^5$, at which the one-bunch
configuration has still survived. These phenomena suggest that the
one-bunch solution is an attractor of the system \cite{q-sol}.

\begin{figure}
\leavevmode
\epsfxsize=8.3cm
\setlength{\unitlength}{1cm}
\begin{picture}(8,6.8)
\put(-0.5,0.2){\epsfbox{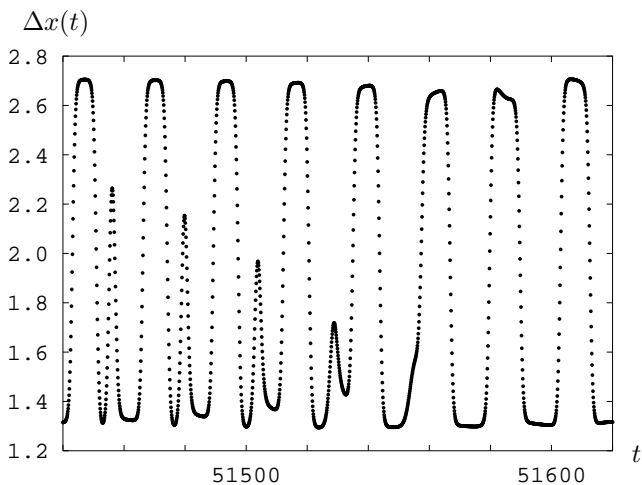}} 
\put(7.9,0.55){$t$}
\put(-0.2,6.3){$\Delta x(t)$}
\end{picture}
\caption{The fusion process at $t\simeq51560$. A bunch starting to
shrink at $t\simeq 48000$ is absorbed by the other}
\end{figure}


\section{Summary and Discussions}
We find a finite set of exact multi-bunch solutions for a
car-following model with driver's reaction time lag in a circuit,
which is described by a differential-difference equation under the
cyclic boundary condition. When the circuit length and the total
number of cars are given, we can calculate the possible number of
car-bunches in the circuit and the profile of the density wave which
describes the car-bunching. The numerical simulation shows that the
linearly unstable uniform flow develops into the one-bunch solution,
which seems to be an attractor of the system. In the relaxation
process from the uniform flow to the one-bunch solution, the system
stays for a while on the configurations which are similar to the
higher multi-bunch solutions. For smaller number of bunches, the
durations of these quasi stable configurations become longer
exponentially (See Fig. 1 in \cite{q-sol}). 

The series of cascade transitions among the multi-bunch configurations
suggests that the multi-bunch solutions correspond to heteroclinic
points of the system. However, any flow out of the one-bunch solution 
does not observed up to the present. As the other possibility, each of 
the multi-bunch solution may be a kind of the Milnor attractor
\cite{milnor}, which is unstable for any small perturbations, but
globally attracts the orbits. To see the problem, we are going to have
to investigate the stability and the attracting domain of the
multi-bunch solutions more precisely.

The density wave formation from a uniform flow on the OV model is very
similar to the present model. The car-bunching and the fusion of
bunches are observed also in the OV model. It may be possible to
expect that the existence of the multi-bunch solutions and the
qualitative feature of the cascade transitions are common to both
models, although the analytic solutions for the OV model have not been
obtained.


\acknowledgements
The author express great appreciation to Y.~Igarashi and K.~Itoh for
many useful discussions. I also thank them for encouraging this work
and carefully reading the manuscript. Thanks due to Y.~Ohno and
T.~Shibata for helpful suggestions and comments. The hospitality of
E-laboratory of Nagoya University is also acknowledged.


\appendix
\section{Definitions and formula of elliptic functions}
The definitions of elliptic functions and some mathematical formula
used in this paper are listed. The elliptic theta functions are
defined as the infinite products;
\begin{eqnarray}
\vth_0(v,q)&=& q_{0}\prod_{n=1}^{\infty}
(1-2q^{2n-1}\cos2\pi v+q^{4n-2}),\nn\\
\vth_1(v,q)&=&2q^{1/4} q_{0}\sin\pi v\prod_{n=1}^{\infty}
(1-2q^{2n}\cos2\pi v+q^{4n}),\nn\\
&&~~{\rm where}\qquad q_{0}=\prod_{n=1}^{\infty}(1-q^{2n}).
\label{a:def_theta}
\end{eqnarray}
The addition formula of the theta functions are
\begin{eqnarray}
\vth_0(v+w)\vth_0(v-w)\vth_0^2(0)
&=&\vth_0^2(v)\vth_0^2(w)-\vth_1^2(v)\vth_1^2(w),\nn\\
\vth_1(v+w)\vth_1(v-w)\vth_0^2(0)
&=&\vth_1^2(v)\vth_0^2(w)-\vth_0^2(v)\vth_1^2(w).\nn\\
\label{a:addition}
\end{eqnarray}
Jacobi's elliptic functions and their modulus $k$ and complementary
modulus $k'$ can be expressed by the theta functions:
\begin{eqnarray}
&&k=\frac{\vth_2^2(0,q)}{\vth_3^2(0,q)},\quad
k'=\frac{\vth_0^2(0,q)}{\vth_3^2(0,q)},\quad
K=\frac{\pi}{2}~\vth_3^2(0,q),\nn\\
&&{\rm sn}2Kv=\frac{1}{\sqrt{k}}~\frac{\vth_1(v)}{\vth_0(v)},\quad
{\rm cn}2Kv=\sqrt{1-{\rm sn}^22Kv},\label{a:jacobi}\\
&&{\rm dn}2Kv=\sqrt{1-k^2{\rm sn}^22Kv},\nn
\end{eqnarray}
where $K$ is the complete elliptic integral of the first kind.


\section{Theta function formalism}
Let us solve the equation of motion,
\begin{equation}
{\dot x}_n(t+\tau)=\xi+\eta~\tanh
\left[\frac{\Delta x_n(t)-\rho}{2\sigma}\right],
\label{a:EoM}
\end{equation}
by the functional ansatz,
\begin{equation}
x_n(t)=Ct-n h + A~{\ln} \frac
{\vth_0\left(v-\beta+\delta\right)}{\vth_0\left(v-\beta-\delta\right)},
\label{a:xn}
\end{equation}
using the variable notations,
\begin{equation}
v\equiv\nu t-\frac{n}{\lambda},\qquad 
\beta\equiv\frac{1}{2\lambda}.
\label{a:v&beta}
\end{equation}
By virtue of the Whitham condition $\nu\tau=\beta$, the velocity with
time lag  $\tau$, ${\dot x}_n(t+\tau)$, can be written as
\begin{equation}
{\dot x}_n(t+\tau)=
C+A\nu\frac{d}{dv}\ln\frac{\vth_0(v+\delta)}{\vth_0(v-\delta)},
\label{a:velo1}
\end{equation}
where we replace the time derivative $d/dt$ with $\nu d/dv$. 
Converting the $v$-derivative into $d/d\delta$ again, we can apply the
addition formula \bref{a:addition} to the expression; 
\begin{eqnarray}
\frac{d}{dv}\ln\frac{\vth_0(v+\delta)}{\vth_0(v-\delta)}&=&
\frac{d}{d\delta}\ln\left[ \vth_0(v+\delta)\vth_0(v-\delta)\right]\nn\\
&=&\frac{d}{d\delta}\ln\left[\vth_0^2(v)\vth_0^2(\delta)
-\vth_1^2(v)\vth_1^2(\delta)\right].\nn\\
\end{eqnarray}
It allows the velocity to be a rational expression of $\vth_0^2(v)$
and $\vth_1^2(v)$;
\begin{equation}
{\dot x}_n(t+\tau)=C+A\nu\frac{\vth_0^2(v)\left(\vth_0^2(\delta)\right)'
-\vth_1^2(v)\left(\vth_1^2(\delta)\right)'}
{\vth_0^2(v)~\vth_0^2(\delta)-\vth_1^2(v)~\vth_1^2(\delta)}.
\label{a:velo2}
\end{equation}

We find that the headway,
\begin{equation}
\Delta x_n(t)=h+A~\ln\frac
{\vth_0(v+\beta+\delta)~\vth_0(v-\beta-\delta)}
{\vth_0(v+\beta-\delta)~\vth_0(v-\beta+\delta)},
\label{a:head1}
\end{equation}
can also be expressed as a rational form of $\vth_0^2(v)$ and
$\vth_1^2(v)$;
\begin{equation}
e^{2X_n}=\frac
{\vth_0^2(v)~\vth_0^2(\delta+\beta)-\vth_1^2(v)~\vth_1^2(\delta+\beta)}
{\vth_0^2(v)~\vth_0^2(\delta-\beta)-\vth_1^2(v)~\vth_1^2(\delta-\beta)},
\label{a:head2}
\end{equation}
where
\begin{equation}
X_n=\frac{\Delta x_n(t)-h}{2A}.
\end{equation}
By eliminating $\vth_0^2(v)$ and $\vth_1^2(v)$ from \bref{a:velo2} and 
\bref{a:head2}, it will be shown that the velocity ${\dot x}_n(t+\tau)$ is
equal to a first order rational expression of $e^{2X_n}$, which can be
rewritten as an hyperbolic tangent function of the headway $\Delta x_n(t)$.

Performing the elimination, we get a differential-difference equation
which the ansatz \bref{a:xn} satisfies;
\begin{equation}
{\dot x}_n(t+\tau)=C+A\nu\frac{N_++N_-e^{2X_n}}{D_++D_-e^{2X_n}},
\end{equation}
where
\begin{eqnarray}
D_{\pm}&=&\vth_0^2(0)\vth_1(\beta)\vth_1(2\delta\pm\beta),\nn\\
N_{\pm}&=&\pm\left[\vth_1^2(\delta\pm\beta)(\vth_0^2(\delta))'
-\vth_0^2(\delta\pm\beta)(\vth_1^2(\delta))'\right].
\end{eqnarray}
We can transform $N_{\pm}$ into the expressions in terms of
$\beta$-derivatives, instead of $\delta$ ones;
\begin{eqnarray}
N_{\pm}&=&\left(\pm\frac{d}{d\delta}-\frac{d}{d\beta}\right)
\left[\vth_1^2(\delta\pm\beta)\vth_0^2(\delta)
-\vth_0^2(\delta\pm\beta)\vth_1^2(\delta)\right]\nn\\
&=&\pm\vth_0^2(0)\vth_1(\beta)\vth_1(2\delta\pm\beta)
\frac{d}{d\beta}\ln\frac{\vth_1(2\delta\pm\beta)}{\vth_1(\beta)}.
\end{eqnarray}
As mentioned above, the differential-difference equation can be
written by using the hyperbolic tangent function as
\begin{eqnarray}
{\dot x}_n(t&+&\tau)=
C +\frac{A\nu}{2}\left(\frac{N_-}{D_-}+\frac{N_+}{D_+}\right)\nn\\
&+&\frac{A\nu}{2}\left(\frac{N_-}{D_-}-\frac{N_+}{D_+}\right)
\tanh\left(X_n+\frac{1}{2}\ln\frac{D_-}{D_+}\right).
\label{a:EoM2}
\end{eqnarray}
Thus, we find the equations to which the ansatz parameters are
subject, by comparing the above equation with the equation of motion
\bref{a:EoM}:
\begin{eqnarray}
\xi&=&C+\frac{A\nu}{2}\frac{d}{d\beta}
\ln\frac{\vth_1(2\delta+\beta)}{\vth_1(2\delta-\beta)},
\label{a:xi1}\\
\eta&=&\frac{A\nu}{2}\frac{d}{d\beta}\ln
\frac{\vth_1^2(\beta)}{\vth_1(2\delta+\beta)\vth_1(2\delta-\beta)},
\label{a:eta1}\\
\rho&=&h-A\ln\frac{\vth_1(2\delta-\beta)}{\vth_1(2\delta+\beta)},
\label{a:rho}\\
\sigma&=&A.
\label{a:sigma}
\end{eqnarray}
In the first two equations, we can decompose the variables $\delta$,
$\beta$ in the theta functions by the addition formula;
\begin{eqnarray}
\xi&=&C+\frac{A\nu}{2}\frac{d}{d\delta}\ln\vth_0(2\delta)\nn\\
&&+\frac{A\nu}{4}\frac{d}{d\delta}\ln
\left[\frac{\vth_1^2(2\delta)}{\vth_0^2(2\delta)}-
\frac{\vth_1^2(\beta)}{\vth_0^2(\beta)}\right],
\label{a:xi2}\\
\eta&=&-\frac{A\nu}{2}\frac{d}{d\beta}\ln
\left[\frac{\vth_0^2(\beta)}{\vth_1^2(\beta)}-
\frac{\vth_0^2(2\delta)}{\vth_1^2(2\delta)}\right].
\label{a:eta2}
\end{eqnarray}
For the first one, $d/d\beta$ is converted into $d/d\delta$, to apply
the addition formula, before the decomposition. All of the equations
\bref{a:xi1}-\bref{a:sigma} can be shown that they are equivalent to the
ones which we found in the previous paper\cite{IIN}, although they are
quite different in the appearances.


\end{document}